\documentclass[sigconf]{acmart}
\AtBeginDocument{%
  \providecommand\BibTeX{{%
    \normalfont B\kern-0.5em{\scshape i\kern-0.25em b}\kern-0.8em\TeX}}}

\usepackage{graphicx}

\usepackage{amsmath}
\usepackage{amsthm}
\usepackage{algorithm}
\usepackage{algorithmic}
\usepackage{graphics}
\usepackage{placeins}
\usepackage{stfloats}
\usepackage{kantlipsum}
\usepackage{tikz}
\usepackage{pgfplots}
\usepackage{multirow}
\usepackage{caption}
\usepackage{amsthm}
\usepackage{textcomp}
\usepackage{adjustbox}
\usepackage{mathtools}  
\usepackage{amsmath}
\usepackage{tabulary}

\usepackage{booktabs}
\usepackage[normalem]{ulem}
\useunder{\uline}{\ul}{}

\usepackage{enumitem}% http://ctan.org/pkg/enumitem
\usepackage{wrapfig}

\usepackage{graphicx}% http://ctan.org/pkg/graphicx
\usepackage{caption,subcaption}% http://ctan.org/pkg/{caption,subcaption}
\newtheorem{definition}{Definition}
\usepackage{fancyhdr}

\let\oldmaketitle\maketitle
\renewcommand{\maketitle}{%
  \oldmaketitle%
  \thispagestyle{fancy}}
\begin{document}
\fancyhead{}
% \pagestyle{fancy}
% \fancyhf{}
% \rhead{CIKM ’20, October 19–23, 2020}
% \lhead{Session: Long - Recommendation System}
% \cfoot{\thepage}
% \pagestyle{myheadings}
% \markright{John Smith\hfill On page styles\hfill}
\title{STP-UDGAT: Spatial-Temporal-Preference User Dimensional Graph Attention Network for Next POI Recommendation}


%========================multi line authors=======

\author{Nicholas Lim}
\affiliation{GrabTaxi Holdings, Singapore}
\email{nic.lim@grab.com}

\author{Bryan Hooi}
\affiliation{Grab-NUS AI Lab, National University of Singapore, Singapore}
\email{dcsbhk@nus.edu.sg}

\author{See-Kiong Ng}
\affiliation{Grab-NUS AI Lab, National University of Singapore, Singapore}
\email{seekiong@nus.edu.sg}

\author{Xueou Wang}
\affiliation{Grab-NUS AI Lab, National University of Singapore, Singapore}
\email{idswx@nus.edu.sg}

\author{Yong Liang Goh}
\affiliation{GrabTaxi Holdings, Singapore}
\email{yongliang.goh@grab.com}

\author{Renrong Weng}
\affiliation{GrabTaxi Holdings, Singapore}
\email{renrong.weng@grab.com}

\author{Jagannadan Varadarajan}
\affiliation{GrabTaxi Holdings, Singapore}
\email{jagan.varadarajan@grab.com}
%========================multi line authors=======

\begin{abstract}
Next Point-of-Interest (POI) recommendation is a longstanding problem across the domains of Location-Based Social Networks (LBSN) and transportation. Recent Recurrent Neural Network (RNN) based approaches learn POI-POI relationships in a local view based on independent user visit sequences. This limits the model's ability to directly connect and learn across users in a global view to recommend semantically trained POIs. In this work, we propose a Spatial-Temporal-Preference User Dimensional Graph Attention Network (STP-UDGAT), a novel explore-exploit model that concurrently exploits personalized user preferences and explores new POIs in global spatial-temporal-preference (STP) neighbourhoods, while allowing users to selectively learn from other users. In addition, we propose random walks as a masked self-attention option to leverage the STP graphs' structures and find new higher-order POI neighbours during exploration. Experimental results on six real-world datasets show that our model significantly outperforms baseline and state-of-the-art methods.
\end{abstract}





\maketitle

\raggedbottom

\section{Introduction}
With the increasing interest  to provide personalized services, service providers such as Location-Based Social Networks (LBSN)  are keen to understand their users better in order to do well on recommendation tasks. Next Point-of-Interest (POI) recommendation has been a longstanding problem for LBSNs to recommend places of interest to their users.  Recently, next POI recommendation has also been found to be important for other applications.  For instance, ride-hailing services are interested to use next POI recommendation to predict the next pick-up or drop-off points of their customers \cite{grabPOI}. In the terrorism domain, it can be used to predict the likelihood of the next state or city POI prone to be attacked by a terrorist group \cite{strnn}.

Next POI recommendation is a challenging task due to its non-linear patterns in user preferences. Early works explored conventional collaborative filtering and sequential approaches such as Matrix Factorisation (MF) and Markov Chains (MC) respectively.  For example, \cite{FPMC-LR} extended the Factorizing Personalized Markov Chain (FPMC) approach \cite{FPMC} that integrates both MF and MC, to include localised region constraints and recommend nearby POIs for the next POI recommendation task.

Recently, several works have proposed Recurrent Neural Network (RNN) based approaches to better model the sequential dependencies of users' historical POI visits to learn their preferences, as well as incorporating spatial and temporal factors in different ways. \cite{strnn} proposed a Spatial Temporal Recurrent Neural Network (ST-RNN) to leverage spatial and temporal intervals between neighbouring POIs,  setting a time window to take several POIs as input. \cite{hstlstm} proposed the Hierarchical Spatial-Temporal Long-Short Term Memory (HST-LSTM) to incorporate spatial and temporal intervals directly into LSTM's existing multiplicative gates. \cite{stgcn} proposed the Spatio-Temporal Gated Coupled Network (STGCN) to capture short and long-term user preference with new time and distance specific gates. \cite{LSTPM} proposed the Long- and Short-Term Preference Modeling (LSTPM) to learn long and short term user preferences through the use of a nonlocal network and a geo-dilated RNN respectively.

With a clear trend towards learning user preferences from these RNN-based approaches, a notable limitation is in how they learn POI-POI relationships with a local view, where a POI is similar to another POI if they tend to co-occur within individual users' visit sequences. This limits the model's ability to directly learn POI-POI relationships across all users in a global view through inter-user POI-POI connections. For example, as shown in Fig. 1, similar users with a common preference in shopping mall POIs can be used to support recommendations to a user who likes shopping malls. This inter-user  preference-based relationship can be leveraged for next POI recommendation. Similarly,  global spatial and temporal factors across users, such as semantically similar POI pairs across users that have small spatial intervals (i.e. nearby) and small temporal intervals (i.e. visited in similar timings), can be useful for learning POI-POI relationships. 

To learn the underlying POI-POI relationships from both local and global views, we propose a Spatial-Temporal-Preference User Dimensional Graph Attention Network (STP-UDGAT), an explore-exploit model for the next POI recommendation task based on Graph Attention Networks (GAT) \cite{gat}. STP-UDGAT learns POI-POI relationships based on spatial, temporal and preference factors by concurrently exploiting personalized user preference neighbourhoods and exploring new global spatial-temporal-preference (STP) neighbourhoods with self-attention. Additionally, STP-UDGAT also learns user-user relationships, allowing users to selectively learn from other similar users. To recommend a POI for a user, the model takes advantage of both local and global neighbourhoods by balancing the explore and exploit trade-offs. For the exploration phase,  we also propose a novel random walk masked self-attention option to traverse the graph structure and selectively attend to relevant higher-order neighbours so that the model does not only focus on first-order neighbours.

To summarise, the following are the contributions of this paper:

\begin{itemize}[leftmargin=*]

  \item We propose a novel STP-UDGAT model to learn POI-POI relationships from both local (i.e. only user herself) and global (i.e. all users) views based on spatial, temporal and preference factors by balancing the explore-exploit trade-offs. STP-UDGAT also learns user-user relationships to support the recommendation task.
  \item We propose a new masked self-attention option of random walks that can leverage the graph structure to identify and attend higher-order neighbours as compared to just first-order neighbours in GAT.
  \item Experiments conducted on six real-world datasets across the domains of LBSN, terrorism and transportation show that our approach outperforms baseline and state-of-the-art methods. To the best of our knowledge, this is the first work to study GAT and how it can be utilized for the next POI recommendation task.

\end{itemize}

\begin{figure}  [t]
  \centering
  \includegraphics[width=5cm,height=2cm]{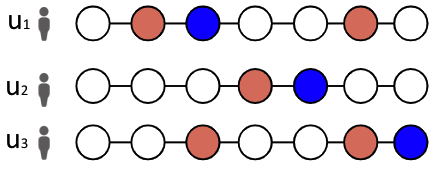}
    \caption{Common shopping mall POIs (red and blue vertices) among different users' visit sequences with the same shopping mall preference.}
\end{figure}
\raggedbottom

\section{Related Work}
\paragraph{\textbf{Next POI Recommendation Task}} We are interested in recommending a ranked set of POIs for a user where the next POI to be visited by the user would be highly ranked. \cite{FPMC-LR} proposed FPMC-LR for the next POI recommendation task by focusing on localised region constraints and exploiting a personalised MC for each user's visit sequence. \cite{PRME} proposed PRME to jointly consider user preference and sequential transitions by modelling POIs and users in a latent space. A Bayesian personalized ranking method \cite{thirdRankTensor} fuses visit behaviours and latent preference of users by considering categorical information. \cite{CAPE} proposed to learn content-aware POI embeddings through user visit sequences and POI textual information. Recently, RNN based approaches have been  proposed to better model the sequential dependencies in the user visit sequences. ST-RNN \cite{strnn} was an early work which showed that spatial and temporal intervals between neighbouring POIs can be utilised in an RNN. To handle the continuous nature of the intervals, ST-RNN performs linear interpolation and learns time and distance specific transition matrices. \cite{hstlstm} proposed ST-LSTM to incorporate spatial and temporal intervals into LSTM's existing multiplicative gates after performing linear interpolation and included a hierarchical variant for session data. \cite{stgcn} proposed STGN, a LSTM based model by introducing dedicated time and distance gates, as well as a separate cell state with the goal to learn both short and long term user preferences. Their variant STGCN, was also proposed to reduce parameters by coupling input and forget gates. A category-aware deep model \cite{categoryAware} includes geographical proximity and POI categories to reduce data sparsity but only predicts the POIs visited in the next 24 hours. \cite{LSTPM} proposed the state-of-the-art LSTPM model to learn long and short term user preferences in a context-aware nonlocal network
architecture that considers the temporal and spatial correlations
between past and current trajectories. This allows the individual learning of the long term preferences (i.e. past trajectories) and the short term preferences (i.e. most recent trajectory) with a nonlocal network and a geo-dilated RNN respectively, before combining them for the recommendation task of the next POI.

\paragraph{\textbf{Graph Representation Learning}} Recently, graph-based methods has been found to be effective in other recommendation problems, such as \cite{sessionBased} which proposed a graph neural network method for session-based recommendation by considering global preference and local factors of session's interest. More recently, motivated by the success of self-attention mechanisms in the Transformer model \cite{transformer}, GAT was introduced to perform masked self-attention and are also effective in recommendation problems. For example, \cite{KGAT} extended GAT for item recommendation by modelling relationships in knowledge graphs. For the next POI recommendation task, our STP-UDGAT is the first work to study GAT, incorporating STP factors to learn both POI-POI and user-user relationships, and using a new masked self-attention option of random walks to attend to higher-order neighbours.

In the works of \cite{adversarial,ensemble,unifying}, they have shown the use of POI-POI graphs to be helpful for learning POI semantics for other predictive tasks. More related to our STP-UDGAT model is an early work of GE \cite{GE} due to its usage of graphs. GE uses a POI-POI graph and bipartite graphs of POI-Region, POI-Time and POI-Word to learn node embeddings, then performs linear combinations of these embeddings in its scoring function to output recommendations. Our proposed STP-UDGAT has several key differences. First, STP-UDGAT's focus is on learning graph representations through GAT and the masked self-attention process, whereas GE focuses on learning node embeddings with LINE \cite{LINE}; both methods have clear differences in algorithm and optimization objectives. Second, our POI-POI and User-User graphs are designed for use by GAT and are not bipartite as bipartite graphs proposed in GE cannot be used by GAT due to the different node types. Third, only STP-UDGAT proposes to learn the balance of explore-exploit trade-offs among the local (i.e. only user herself) and global (i.e. all users) views. Additionally, STGCN \cite{stgcn} has showed GE to perform significantly poorer as compared to basic recurrent baselines of RNN, GRU and LSTM on all of their datasets for all metrics in their work, whereas our STP-UDGAT does not just surpass these recurrent baselines, but also the state-of-the-art LSTPM significantly.

\section{Preliminaries}

\paragraph{\textbf{Problem Formulation}} Let  $U=\{u_{1},u_{2},...,u_{M}\}$ be a set of $M$ users and $V=\{v_{1},v_{2},...,v_{N}\}$
be a set of $N$ POIs for the users in $U$ to visit. Each user $u_{m}$ has a sequence of POI visits $s_{u_{m}}=\{v_{t_{1}},v_{t_{2}},...,v_{t_{i}}\}$ and $S$ is the set of visit sequences for all users where $S=\{s_{u_{1}},s_{u_{2}},...,s_{u_{M}}\}$. The objective of the next POI recommendation task is to consider the historical POI visits $\{v_{t_{1}},v_{t_{2}},...,v_{t_{i-1}}\}$ and user $u_{m}$ to recommend an ordered set of POIs from $V$, where the next POI visit $v_{t_{i}}$ should be highly ranked in the recommendation set. We further denote $V^{train}$, $s_{u_{m}}^{train}$ and $S^{train}$ as sets from the training partition.

\paragraph{\textbf{GAT}} \cite{gat} follows the ``masked'' self-attention process (i.e. masked to consider only adjacent vertices) to compute a hidden representation for vertex $i$ by attending to each vertex in its neighbourhood set $N[i]$ from a graph $G$. A single head GAT layer $\Phi_{\Theta}$ can be abbreviated as:
\begin{gather}
  \vec y_{i} = \Phi_{\Theta}(\vec{i},\hat{N}_{G}^{A}[i])
\end{gather}
where $\vec y_{i}\in \mathbb{R}^{\delta}$ is the output hidden representation of the GAT layer $\Phi_{\Theta}$ that accepts a tuple of $(\vec{i},\hat{N}_{G}^{A}[i])$,   $\vec{i}\in\mathbb{R}^{d}$ as the input representation of vertex $i$ and $\hat{N}_{G}^{A}[i]$ as the set of $n$ neighbours, where each neighbour $\vec{j} \in \hat{N}_{G}^{A}[i]$ has their own input representation $ \vec{j} \in\mathbb{R}^{d}$. 
In $\hat{N}_{G}^{A}[i]$, the $n$ neighbours are determined from the closed neighbourhood of vertex $i$ based on the adjacency option denoted as $A$ from a graph $G$ (i.e. first-order neighbours and vertex $i$ itself). 
% first project both $\vec{x_{i}}$ and $N_{G}^{A}[i]$ to the same hidden space with $\textbf{W}_{p} \in \mathbb{R}^{d \times \delta}$, then

Given the input tuple $(\vec{i},\hat{N}_{G}^{A}[i])$, a GAT layer first performs the self-attention process by computing scalar attention coefficients $\alpha_{ij}\in \mathbb{R}$ for each neighbouring vertex's representation $\vec{j} \in \hat{N}_{G}^{A}[i]$ in the scale of 0 and 1, where 1 means ``completely attend vertex $j$'' and 0 means ``completely ignore vertex $j$''. This involves the use of an input projection weight matrix $\textbf{W}_{p} \in \mathbb{R}^{d \times \delta}$ and a linear projection $\textbf{a}$ parameterized with $\{\textbf{W}_{a} \in \mathbb{R}^{2\delta}, \textbf{b}_{a} \in \mathbb{R}\}$:
\begin{gather}
\alpha_{ij} = \frac{exp\left(LeakyReLU\left(\textbf{a}\:[\textbf{W}_{p}\vec{i}\:||\:\textbf{W}_{p}\vec{j}]\right)\right)}{\sum_{\vec{k} \in \hat{N}_{G}^{A}[i]} exp\left(LeakyReLU\left(\textbf{a}\:[\textbf{W}_{p}\vec{i}\:||\:\textbf{W}_{p}\vec{k}]\right)\right)}
\end{gather}
where $||$ is the concatenation operation, $LeakyReLU$ as the non-linear activation function and the softmax function to output the attention coefficients as a probability distribution that sums to 1 for all $n$ neighbours. With the learned coefficients, a weighted sum between vertex $i$ and its neighbours in $\hat{N}_{G}^{A}[i]$ is then computed as the output hidden representation of the GAT layer $\Phi_{\Theta}$:
\begin{gather}
\vec y_{i} = \sum_{\vec{j} \in \hat{N}_{G}^{A}[i]} \alpha_{ij}  \textbf{W}_{p} \: \vec{j}
\end{gather}

\section{Approach}
Our approach is to learn POI-POI and user-user relationships from both local (i.e. only user herself) and global (i.e. all users) views based on STP factors. In this section, we first propose the Dimensional GAT (DGAT) to learn attention coefficients across dimensions to improve the self-attention process. Then, we introduce the Personalized-Preference DGAT (PP-DGAT) to exploit each user's historical POI visits or local POI preferences, followed by extending it to Spatial-Temporal-Preference DGAT (STP-DGAT) that not only performs the same exploitation of users' local POI preferences, but also includes the exploration of global STP graphs to consider new POIs which the user has never visited before, as well as balancing the explore-exploit trade-offs among the local (exploit) and global (explore) views. Lastly, we further introduce UDGAT (User-DGAT) to allow users to learn to attend to other similar users.

\subsection{DGAT}
In a GAT layer, the self-attention process first computes scalar attention coefficients with a shared linear projection $\textbf{a}$ for each neighbour $\vec{j} \in \hat{N}_{G}^{A}[i]$ in the scale of 0 and 1, as per Eq. (2). Then, in Eq. (3), the predicted coefficients are used for a weighted sum to compute a hidden representation $\vec y_{i}$ accordingly. This process makes a key assumption where the scalar coefficients are representative of the whole vector representation for each neighbour $\vec{j} \in \hat{N}_{G}^{A}[i]$. We argue that self-attention can be applied to each dimension of $\vec{j}$ to better leverage the latent semantics, where each dimension would have its own coefficient. To extend the scalar attention coefficients (GAT) to dimensional attention coefficients (DGAT), first, we modify the linear projection $\textbf{a}$ to predict $\delta$ dimensional coefficients instead of 1 (i.e. $\{\textbf{W}_{a} \in \mathbb{R}^{2\delta}, \textbf{b}_{a} \in \mathbb{R}\}$ to $\{\textbf{W}_{a} \in \mathbb{R}^{2\delta\times\delta}, \textbf{b}_{a} \in \mathbb{R}^{\delta}\}$), resulting Eq. (2) to output $\vec{\alpha}_{ij}\in \mathbb{R}^{\delta}$ instead of scalar $\alpha_{ij}\in \mathbb{R}$. Secondly, we replace Eq. (3) with:
\begin{gather}
    \vec y_{i} = \sum_{\vec{j} \in \hat{N}_{G}^{A}[i]} \vec{\alpha}_{ij} \odot \textbf{W}_{p} \: \vec{j}
\end{gather}
where $\odot$ is the Hadamard product to achieve the intention of DGAT. 

\subsection{PP-DGAT}

Applying DGAT to the next POI recommendation problem is not a straight forward task. For instance, given the historical POI visits $\{v_{t_{1}},v_{t_{2}},...,v_{t_{i-1}}\}$ for a user $u_{m}$, we would like to predict the next POI $v_{t_{i}}$. We can use the previous POI $v_{t_{i-1}}$ as input to the DGAT layer to output a hidden representation from the masked self-attention process by attending to a set of reliable reference POI neighbours $N_{G}^{A}[v_{t_{i-1}}]$ queried from a graph $G$ given $v_{t_{i-1}}$, however, it is unclear how this neighbourhood and graph can be constructed such that the queried closed neighbourhood  $N_{G}^{A}[v_{t_{i-1}}]$ (i.e. adjacent POI vertices and $v_{t_{i-1}}$ itself) are indeed relevant to vertex $v_{t_{i-1}}$ and can benefit the overall prediction task.

\begin{definition}[personalized preference graph]
An undirected complete POI-POI graph for each user $u_{m}\in U$, denoted as $u_{m}^{G}= (V_{u_{m}},E_{u_{m}})$ where $V_{u_{m}}$ and $E_{u_{m}}$ are the sets of POIs and unweighted edges respectively. We set $V_{u_{m}}$=$s^{train}_{u_{m}}$ and all pairs of POI vertices are connected, forming a complete graph that represents the user's historical POI preferences.
\end{definition}

\paragraph{\textbf{Learning From Local View}} As per Definition 1, we propose to construct a fully connected or complete Personalized Preference (PP) graph $u_{m}^{G}$ from each user's set of historical training POIs $s_{u_{m}}^{train}$, where $u_{m}^{G}$ can be actively queried for $N_{G}^{A}[v_{t_{i-1}}]$ in each time step of the prediction by a DGAT layer. Fig. 2 illustrates how a user's available historical POIs $s_{u_{m}}^{train}$ are used to construct a PP graph that serves to be queried by the DGAT layer for the closed neighbourhood $N_{G}^{A}[v_{t_{i-1}}]$ when given input of $v_{t_{i-1}}$ and performing the self-attention process. Accordingly, this would also mean that each user $u_{m}$ will have her own PP graph $u_{m}^{G}$ that encapsulate her own ``local'' historical preferences without clear consideration of other users' POI visiting behaviors to learn POI-POI relationships. This enables personalization by exploiting only the user's individual preferences from a local view as personalization has been shown by past works (e.g. FPMC-LR \cite{FPMC-LR}) to surpass ``global'' methods that consider all users' behaviors directly (e.g. MC).

Next, we describe the PP-DGAT in detail. Given a previous visit POI $v_{t_{i-1}}$ as input, we query the user's PP graph $u_{m}^{G}$ for $N^{A}_{u_{m}^{G}}[v_{t_{i-1}}]$ as the set of reference POI neighbours and is equivalent to all vertices $V_{u_{m}}$ in $u_{m}^{G}$ due to the completeness design of the PP graph and the closed neighbourhood nature of the query (i.e. including $v_{t_{i-1}}$). This essentially allows the DGAT layer to perform self-attention on all historical training POIs of the user because $V_{u_{m}}$=$s^{train}_{u_{m}}$ by definition. Then, we provide both $v_{t_{i-1}}$ and $N^{A}_{u_{m}^{G}}[v_{t_{i-1}}]$ to $Emb$ as an input tuple, where $Emb$ is an embedding layer, parameterised by the POI weight matrix $\textbf{W}^{poi} \in \mathbb{R}^{|V|\times dim}$ and $dim$ is the defined embedding dimension. Accordingly, $Emb$ outputs the tuple of the corresponding POI embedding representations $(\vec{v}_{{t_{i-1}}},\hat{N}^{A}_{u_{m}^{G}}[v_{t_{i-1}}])$, $\vec{v}_{{t_{i-1}}}$ as the embedding of the previous POI $v_{t_{i-1}}$, and $\hat{N}^{A}_{u_{m}^{G}}[v_{t_{i-1}}]$ as the set of embeddings of all the neighbors of $v_{t_{i-1}}$:
\begin{gather}
  (\vec{v}_{{t_{i-1}}},\hat{N}^{A}_{u_{m}^{G}}[v_{t_{i-1}}]) = Emb_{\textbf{W}^{poi}}(v_{{t_{i-1}}},N^{A}_{u_{m}^{G}}[v_{t_{i-1}}]) \\
  \vec y_{t_{i}} = \Phi_{\textbf{W}^{PP}}(\vec{v}_{{t_{i-1}}},\hat{N}^{A}_{u_{m}^{G}}[v_{t_{i-1}}]) \\
  P(v_{t_{i}}|v_{t_{i-1}}) = softmax(FC_{\textbf{W}^{f1}}(\:D(\vec y_{t_{i}})\:))
\end{gather}  
Using a DGAT layer $\Phi_{\textbf{W}^{PP}}$, Eq. (6) computes the hidden representation $\vec y_{t_{i}}$ and Eq. (7) linearly projects $\vec y_{t_{i}}$ to the number of classes or POIs (i.e. $|V|$) followed by a softmax function where $D$ is the dropout layer, $FC_{\textbf{W}^{f1}\in \mathbb{R}^{\delta \times |V|} }$ as a linear Fully Connected (FC) layer.
% parameterised with $\textbf{W}^{f1} = \{ \textbf{W},  \textbf{b} \: | \: \textbf{W}\in \mathbb{R}^{\delta\times |V|}, \textbf{b} \in \mathbb{R}^{|V|}\}$. 
With the probability distribution of all POIs in $V$ by learning $P(v_{t_{i}}|v_{{t_{i-1}}})$ as a multi-classification problem, we would have the final ranked POI recommendation set by sorting it in descending order. 
At test time, we follow the same as training to perform self-attention on all vertices of the user's PP graph (i.e. all user's available historical POIs).

\begin{figure}  
  \centering
  \includegraphics[width=\linewidth-0.8cm,height=2.9cm]{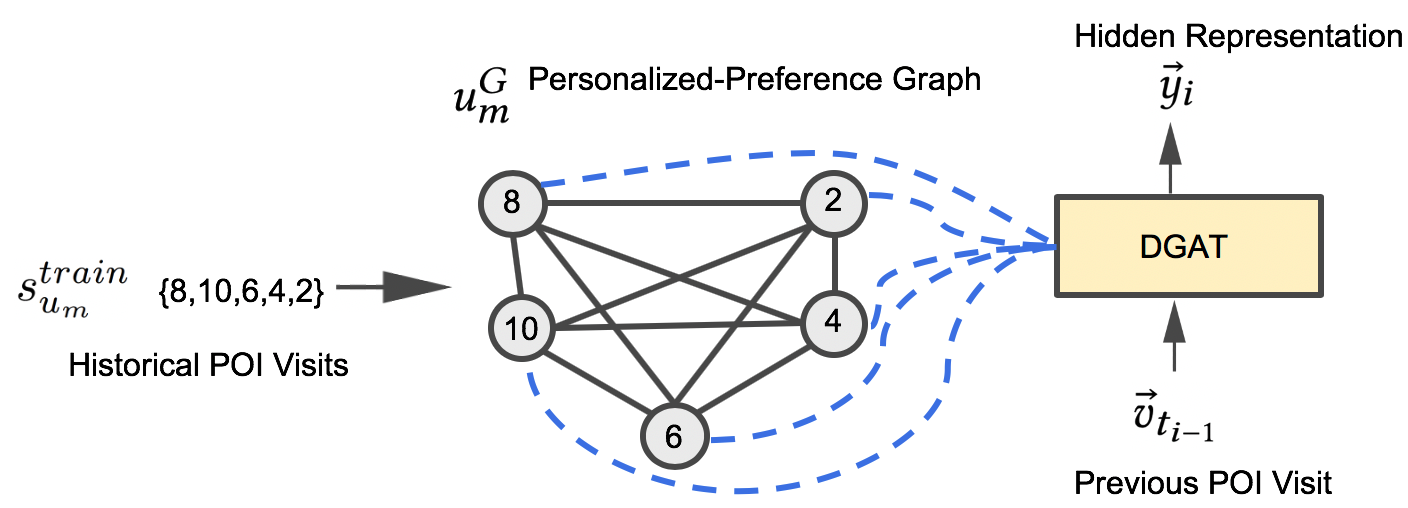}
%  \captionsetup{width=0.8\textwidth}
  \caption{Illustration of self-attention performed on a user's PP graph based on her historical POIs.}
\end{figure}

 \begin{figure*} 
  \centering
  \includegraphics[width=\linewidth-2.4cm,height=4.3cm]{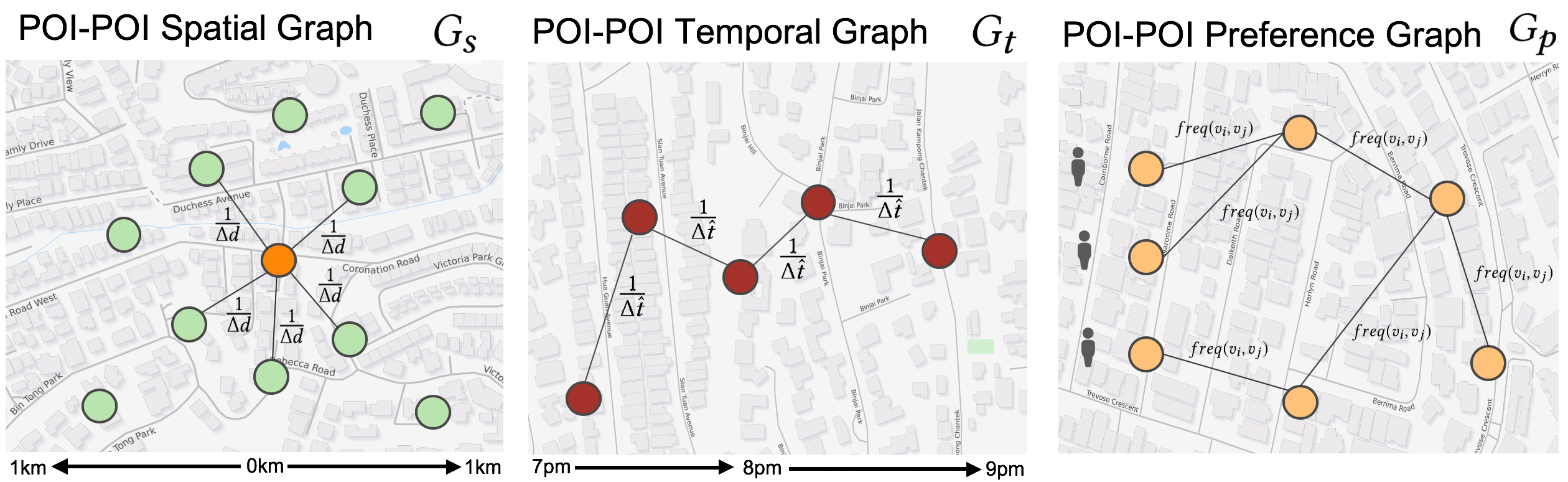}

  \caption{Illustration of how global spatial, temporal and preference factors are represented into POI-POI graphs. Maps © OpenStreetMap contributors, CC BY-SA.}
%   \vspace{-1em}
\end{figure*}

\subsection{STP-DGAT} 
With PP-DGAT, next POI predictions can be computed from just the exploitation of the users' historical POIs $s_{u_{m}}^{train}$ or local preferences. However, this limits the learning of POI-POI relationships for the recommendation task by not considering global spatial, temporal and preference factors. For example, a user who likes shopping mall may be interested in visiting new nearby malls (spatial), or perhaps new malls that are popular only at night (temporal), or new malls which other similar users of the same shopping mall interest have visited (preference). Here, a \emph{new} POI $v_{n}$ refers to an unvisited POI in the user's historical POI visits (i.e. $v_{n} \notin s_{u_{m}}^{train}$). We propose to not only consider local user POI preferences as done in PP-DGAT, but also global STP factors across all users to improve the recommendation task and user experience \cite{treeBased} through the exploration of new unvisited POIs to better learn the POI-POI relationships. Although personalization has been shown by existing works to learn the users' semantics well, we believe that new unvisited POIs identified across global STP factors can also be leveraged in a way that does not jeopardize personalization. We propose to achieve this by learning the balance between the exploitation of the user's local preference or personalization and the exploration of new unvisited POIs based on the global STP factors for the task.

\begin{definition}[Spatial Graph]
An undirected POI-POI graph denoted as $G_{s}= (V_{s},E_{s})$ where $V_{s}$ and $E_{s}$ are the sets of POIs and edges respectively and we set $V_{s}=V$. POI $v_{i}$ has adjacency to POI $v_{j}$ if $v_{j}$ is within the top $\sigma$ nearest POIs based on the distance interval $\Delta d$ using a distance function $d(v_{i},v_{j})$. We set $\sigma=5$ and $d(v_{i},v_{j})$ as Euclidean distance. The edge weight between each pair is $\frac{1}{\Delta d}$.
\end{definition}

\begin{definition}[Temporal Graph]
An undirected POI-POI graph denoted as $\:G_{t}= (V_{t},E_{t})$ where $V_{t}$ and $E_{t}$ are the sets of POIs and edges respectively and $V_{t}=V^{train}$. As each POI visit includes timestamp data, we first combine all users' historical POI visit sequence $s_{u_{m}} \in S^{train}$ to a single set of $S^{train}_{time}$ that disregards which user the POI visits belong to and have all POI visits sorted in chronological order. Then, we compute POI pairs from $S^{train}_{time}$ where POI $v_{i}$ is adjacent to $v_{j}$ if $v_{j}$ is the next visit based on chronological order. The edge weight between each pair is $\frac{1}{\Delta \hat{t}}$ where $\Delta \hat{t}$ is the averaged time interval of all same pairs.
\end{definition}

\begin{definition}[Preference Graph]
An undirected POI-POI graph denoted as $G_{p}= (V_{p},E_{p})$ where $V_{p}$ and $E_{p}$ are the sets of POIs and edges respectively and we set $V_{p}=V^{train}$. POI pairs are computed from each user's visit sequence $s_{u_{m}} \in S^{train}$ where POI $v_{i}$ is adjacent to $v_{j}$ if $v_{j}$ is the next visit. The edge weight between each pair is $freq(v_{i},v_{j})$ where $freq$ is the count function of POI pair occurrences.
\end{definition}

%#see
\paragraph{\textbf{Representing STP Factors}} From Definitions 2 to 4, we propose the spatial, temporal and preference POI-POI graphs to embed the semantics of global STP factors, for the purpose of being utilized by DGAT layers to explore new POIs and leveraging them for the recommendation task. Fig. 3 illustrate examples of how the STP factors are represented into POI-POI graphs and its intentions (whole graph not shown). For instance, in Fig. 3, spatial graph $G_{s}$ connects each POI (orange node) in $V$ to its nearest top 5 POIs to embed geographical proximities; temporal graph $G_{t}$ connects POIs in $S^{train}_{time}$, as per definition, connecting POIs that are similar in visit timestamps across all users and disregarding which user the POI visit it belongs to. This allows schools and bus stations POIs to be connected during day time, and similarly so for bars and clubs POIs at night even when they do not co-occur in a user's historical POI visit sequence.  Preference graph $G_{p}$ connects POIs sequentially across each user's sequence for all users, allowing POI vertices to connect in a global preference view by considering all users and connecting similar sequences, which contrasts with a PP graph that only considers the historical POIs of the user herself.

\paragraph{\textbf{Exploring New STP Neighbours}} With the proposed graphs, we would like to find new POIs from STP graphs which the user has never visited before, but yet are relevant to the user and can help better learn POI-POI relationships. We propose to use all vertices $V_{u_{m}}$ in the user's PP graph $u_{m}^{G}$ as the set of seed POIs to find relevant new POIs in STP graphs. As all vertices $V_{u_{m}}$ is equivalent to the user's historical POIs, $V_{u_{m}}=s_{u_{m}}^{train}$ as per Definition 1, this allows us to find relevant neighbours based on the entire vertex set. First, we compute the closed neighbourhood of each POI in the user's PP graph vertex set  $\{N^{A}_{G}[v_{i}] \: | \: v_{i} \in  V_{u_{m}}\} $ through the adjacency option $A$ from a graph $G$, then, we remove all visited POIs by the user and perform a frequency ranking to compute the top $\tau$ new neighbours, denoting this result set $N^{A}_{G}(V_{u_{m}})^{\tau}$ as the open neighbourhood of $V_{u_{m}}$ (i.e. excluding POIs in $V_{u_{m}}$ to keep only newly discovered POIs). We thus identify $N^{A}_{G}(V_{u_{m}})^{\tau}$ for all proposed STP graphs $G_{STP} = \{G_{s},G_{t},G_{p}\}$ and mapping them to POI representations with Eq. (5). Then, we use separate DGAT layers and perform mean pooling to output a single STP representation $\vec o^{A}_{t_{i}}$: 
\begin{gather}
\vec o^{A}_{t_{i}} = \frac{1}{|G_{STP}|} \:\:\:\:\sum_{\mathclap{\substack{G\in G_{STP} \\ 
  \textbf{W} \in \{\textbf{W}_{S}^{A}, \textbf{W}_{T}^{A} \textbf{W}_{P}^{A} \} }}}\Phi_{\textbf{W}}(\vec{v}_{{t_{i-1}}},\hat{N}^{A}_{G}(V_{u_{m}})^{\tau})
\end{gather}

\begin{figure}
  \centering
  \includegraphics[width=4.5cm,height=3.0cm]{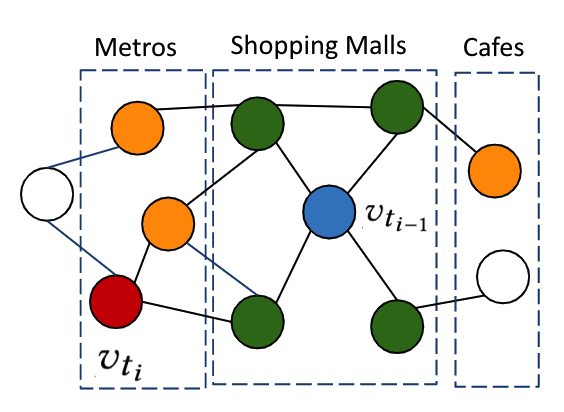}
  \caption{Input previous POI of $v_{t_{i-1}}$ (blue vertex) and next POI of $v_{t_{i}}$ (red vertex). Green vertices are first-order neighbours and orange vertices are higher-order neighbours found with random walks.
  }
%   \vspace{-1em}
\end{figure}

\paragraph{\textbf{Random Walk Masked Self-Attention}}
To ensure that the exploration of new POIs in STP graphs is not limited to just new first-order neighbours using adjacency option $A$, we propose an alternative approach of random walks that have been shown to be effective in exploring diverse neighbourhoods of graphs in other applications \cite{node2vec}, where $N^{RW}_{G}(V_{u_{m}})^{\tau}$ is the top $\tau$ new STP neighbours found with random walk masked self-attention option $RW$ to focus on higher-order neighbours. Fig. 4 illustrates the problem where first-order neighbours are insufficient to represent the neighbourhood of vertex $v_{t_{i-1}}$, a shopping mall, and can benefit from higher-order vertices as determined by the random walks, such as nearby metros to correctly predict $v_{t_{i}}$, a metro. As per Definitions 2 to 4 and Fig. 3, we set the edge weights of STP graphs as $\frac{1}{\Delta d}$, $\frac{1}{\Delta \hat{t}}$ and $freq(v_{i},v_{j})$ respectively. This intentionally biases the random walks to nearby POIs (spatial), POIs visited in similar timings (temporal) and popular POIs visited by other similar users (preference). The total random walk POIs for a graph $G(V,E)$ is computed as $|V|\times \mu \times \beta$ where $\mu$ is the number of random walks per vertex and $\beta$ is the walk's length. 

Effectively, the neighbourhood for a graph $G \in G_{STP}$ can now be computed either through the adjacency option $N^{A}_{G}(V_{u_{m}})^{\tau}$  or the newly proposed random walk option $N^{RW}_{G}(V_{u_{m}})^{\tau}$. Similar to Eq. (8), we compute another STP representation $\vec o^{RW}_{t_{i}}$ based on $RW$ with a separate set of DGAT layers after mapping them to POI representations using Eq. (5):
\begin{gather}
\vec o^{RW}_{t_{i}} = \frac{1}{|G_{STP}|} \:\:\:\:\sum_{\mathclap{\substack{G\in G_{STP} \\ 
  \textbf{W} \in \{\textbf{W}_{S}^{RW}, \textbf{W}_{T}^{RW} \textbf{W}_{P}^{RW} \} }}}\Phi_{\textbf{W}}(\vec{v}_{{t_{i-1}}},\hat{N}^{RW}_{G}(V_{u_{m}})^{\tau})
\end{gather}
Then, to leverage both newly explored first-order (adjacency option $A$) and higher-order (random walks option $RW$) POI neighbours, we fuse both masked self-attention STP representations with a linear layer $FC_{\textbf{W}^{f2}\in \mathbb{R}^{2\delta \times \delta}} $ :
% Using Eq. (5) but with $RW$ and a separate set of DGAT layers $\textbf{W} \in \{\textbf{W}_{S}^{RW}, \textbf{W}_{T}^{RW}, \textbf{W}_{P}^{RW} \} $, we can obtain $\vec o^{RW}_{t_{i}}$ and 
\begin{gather}
  \vec {stp}_{t_{i}} =  FC_{\textbf{W}^{f2}}\:(\:\vec o^{A}_{t_{i}}\:||\:\vec o^{RW}_{t_{i}}\:)
\end{gather}
where $||$ is the concatenate operation and $\vec {stp}_{t_{i}}$ serves as the output hidden representation of the exploration module, based on only newly explored unvisited POIs.

\begin{figure*} 
  \centering
  \includegraphics[width=\linewidth-4.5cm,height=6.0cm]{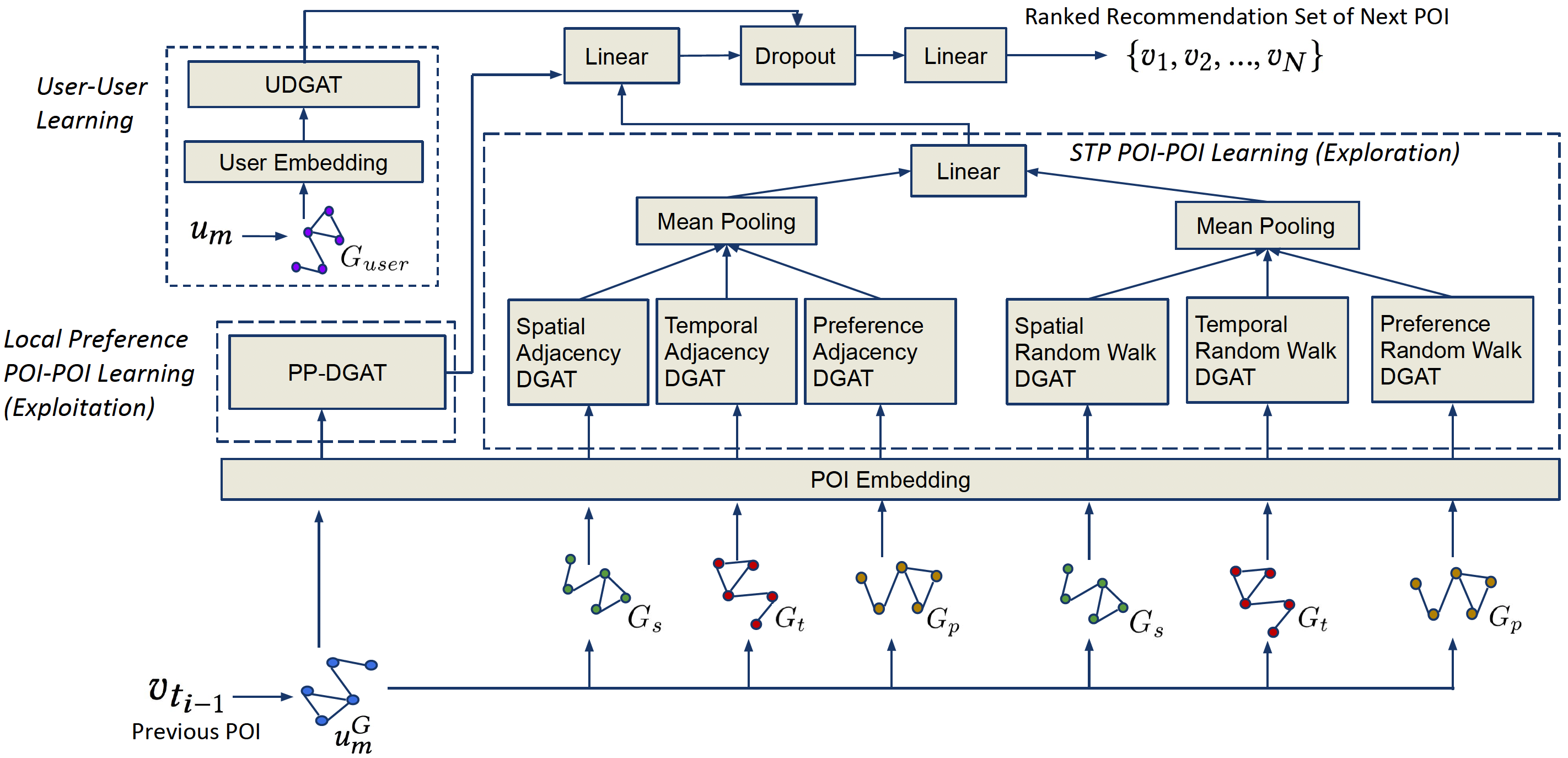}
%  \captionsetup{width=0.8\textwidth}
  \caption{Illustration of STP-UDGAT for the next POI recommendation task.}
\end{figure*}
\paragraph{\textbf{Explore-Exploit}} Next, we propose the use of a linear layer to learn and balance the explore-exploit trade-offs between PP-DGAT's representation (exploiting users' historical POIs) and the STP representation $\vec {stp}_{t_{i}}$ derived from the exploration module as per Eq. (10). Specifically, we update Eq. (6) to Eq. (11) with a new linear layer $FC_{\textbf{W}^{f3}\in \mathbb{R}^{2\delta \times \delta} }$ to fuse and learn the balance between PP-DGAT (exploit) and $\vec {stp}_{t_{i}}$ (explore) concurrently:
\begin{gather}
      \vec y_{t_{i}} = FC_{\textbf{W}^{f3}}(\:\Phi_{\textbf{W}^{PP}}(\vec{v}_{{t_{i-1}}},\hat{N}^{A}_{u_{m}^{G}}[v_{{t_{i-1}}}])\:||\: \vec {stp}_{t_{i}})
\end{gather}

\subsection{STP-UDGAT} ST-RNN \cite{strnn} found the inclusion of user embeddings to be effective; however, this could overfit the model to give only high probabilities to POIs the user has been before. To this end, we propose User DGAT (UDGAT) with the goal of allowing users to learn to attend to other similar users:
\begin{definition}[User Graph]
An undirected user-user graph denoted as $G_{user}= (V_{user},E_{user})$ where $V_{user}$ and $E_{user}$ are the sets of users and edges respectively and we set $V_{user}=U$. User $v_{i}$ has adjacency to user $v_{j}$ if their Jaccard similarity coefficient is above 0.2 (i.e. $\frac{| s^{train}_{v_{i}} \bigcap s^{train} _{v_{j}} |} {| s^{train}_{v_{i}} \bigcup s^{train}  _{v_{j}} |} > 0.2$).
\end{definition}

Given $u_{m}$, we first map the tuple of $u_{m}$ and it's user neighbours $N^{A}_{G_{user}}[u_{m}]$ to its corresponding embedding representations with another embedding layer $Emb$ parameterised by the user weight matrix $\textbf{W}^{user} \in \mathbb{R}^{|U|\times dim}$:
\begin{gather}
  (\vec{u}_{m},\hat{N}^{A}_{G_{user}}[u_{m}]) = Emb_{\textbf{W}^{user}}(u_{m},N^{A}_{G_{user}}[u_{m}]) \\
\vec u_{t_{i}} = \Phi_{\textbf{W}^{u}}(\vec{u}_{m},\hat{N}^{A}_{G_{user}}[u_{m}]) \\
  P(v_{t_{i}}|v_{t_{i-1}}) = softmax(FC_{\textbf{W}^{f1}}(\: D(\vec y_{t_{i}}\:||\:\vec u_{t_{i}})\:))
\end{gather}
Then in Eq. (13), we use a DGAT layer to compute the hidden representation $\vec u_{t_{i}}$ of user $ u_{m}$ by attending to other similar users' embeddings and the user herself. Lastly, we update Eq. (7) to Eq. (14) to fuse STP-DGAT's output hidden representation $\vec y_{t_{i}}$ with $\vec u_{t_{i}}$ to include user semantics through concatenation and updating the linear layer to $FC_{\textbf{W}^{f1}\in \mathbb{R}^{2\delta \times |V|} }$, leading to the final variant of STP-UDGAT as illustrated in Fig. 5.

% through querying for its neighbourhood $N^{A}_{G_{user}}[u_{m}]$ from the proposed user graph with the adjacency option $A$.
 
\section{Experiments}

\subsection{Datasets}

We use six real world datasets across the three domains of:

\begin{itemize}[leftmargin=*]
\item \textbf{LBSN}:
Foursquare Global Scale (Foursquare-Global), Foursquare Singapore (Foursquare-SG),
Gowalla and Brightkite \cite{gowallaBrightkiteData,globalScaleData,sgFS} are well-known public social media datasets used by many works to evaluate for the next POI recommendation task. The data consist of sequential POI check-ins with the goal to predict the next POI.

\item \textbf{Terrorism}: Global Terrorism Database (GTD) \cite{gtd} consists of around 190,000 terrorism incidents across the world since 1970 and is publicly available. Similar to \cite{strnn}, we apply the next POI recommendation task to GTD with the goal of predicting the likelihoods of the city POIs prone to be attacked next by the terrorists based on their historically attacked city POIs so that early preventive actions can be taken.

%% check
\item \textbf{Transportation}: Different from taxi trajectory datasets \cite{MPE} that record taxi-visited POIs from multiple different customers for the same taxi, user trajectory datasets instead record the taxi riding patterns from the same user, and are privately available to ride-hailing companies such as Uber, Didi, and others through the use of their mobile applications. Here, we use a user trajectory dataset of a Southeast Asia (SEA) country (Transport-SEA) from the ride-hailing company Grab to predict the next drop-off point POI based on the user's historical drop-off point POIs.

\end{itemize}

Table 1 shows the details of the datasets.  For preprocessing, we group the datasets into Large and Small Scale categories and only keep POIs visited by more than 10 users for all datasets. We keep users with visit counts between 10 and 30 for Large Scale datasets, and between 10 and 150 for Small Scale datasets. For only the Foursquare-Global dataset, we take the top 40 popular countries by the number of visits. Lastly, we sort each user's visit records by timestamps in ascending order, taking the first 70\% as training set and remaining 30\% as testing set.

\subsection{Baseline Methods and Evaluation Metrics}

\begin{itemize}[leftmargin=*]
\item \textbf{TOP} and \textbf{U-TOP}: These rank POIs using frequencies across $S^{train}$ and in $s_{u_{m}}^{train}$ respectively.
\item \textbf{MF} \cite{MF}: MF is a popular classical approach to many recommendation problems.

  \item \textbf{RNN} \cite{rnn}: RNN takes advantage of sequential dependencies in POI visit sequences with a basic recurrent structure. \textbf{LSTM} \cite{lstm} and \textbf{GRU} \cite{gru} are variants of RNN with different multiplicative gates.

  \item \textbf{HST-LSTM} \cite{hstlstm}: This method incorporates spatial and temporal intervals into LSTM gates. Same as \cite{stgcn}, we use the ST-LSTM variant here as the data does not include session information.
  \item \textbf{STGN} \cite{stgcn}: An LSTM variant that models both short and long term POI visit preferences with new time and distance gates, and cell state. The coupled gate variant \textbf{STGCN} removes the forget gate for better efficiency.
  \item \textbf{LSTPM} \cite{LSTPM}: An LSTM-based model that captures long term preferences with a nonlocal network and short term preferences with a geo-dilated network. LSTPM is the state-of-the-art method for the next POI recommendation task.

\end{itemize}

For our proposed model, we evaluate with the following variants:

\begin{itemize}[leftmargin=*]

\item  \textbf{PP-DGAT-Skip}: Our proposed PP-DGAT model but with an addition of a skip connection to learn the residual function $f(x) + x$ where $f(.)$ is the PP-DGAT model $\Phi_{\textbf{W}^{PP}}$ and $x$ is the previous POI input $\vec{v}_{{t_{i-1}}}$. Specifically, just for this variant, we extend Eq. (6) to:
\begin{gather*}
  \vec y_{t_{i}} = \Phi_{\textbf{W}^{PP}}(\vec{v}_{{t_{i-1}}},\hat{N}^{A}_{u_{m}^{G}}[v_{t_{i-1}}])\:+\: \vec{v}_{{t_{i-1}}}
\end{gather*}

\item \textbf{STP-DGAT}: Our proposed explore-exploit variant that performs exploitation of user's local personalized preferences with PP-DGAT (without skip connection) and exploration of new unvisited POIs in STP graphs and neighbourhoods using both masked self-attention options of $A$ and $RW$.

\item \textbf{STP-UDGAT}: Our final variant of STP-DGAT with UDGAT to include user semantics and allowing users to learn from other users.

\end{itemize}

% \usepackage{booktabs}
% \usepackage{multirow}
\begin{table}[]
\centering
\caption{Statistics of the six datasets (after preprocessing).}
\resizebox{\linewidth}{1.7cm}{%
\begin{tabular}{@{}llllll@{}}
\toprule
\textbf{Categories} & \textbf{Domain} & \textbf{Dataset} & \textbf{\#User} & \textbf{\#POI} & \textbf{\#Visits} \\ \midrule %#see, change to country.
\multirow{5}{*}{Large Scale} & \multirow{1}{*}{Transport} 
  & Transport-SEA & 6,579 & 1,561 & 69,823 \\ \cmidrule(l){2-6} 
 & \multirow{5}{*}{LBSN} & Gowalla\textsuperscript{1} & 9,015 & 2,110 & 51,391 \\
 &  & Brightkite\textsuperscript{2} & 2,377 & 215 & 21,127 \\
 &  & Foursquare-Global\textsuperscript{3} & 10,587 & 1,937 & 64,265 \\ \cmidrule(r){1-1} \cmidrule(l){3-6}
\multirow{2}{*}{Small Scale} &  & Foursquare-SG\textsuperscript{5} & 1,670 & 1,310 & 60,354 \\ \cmidrule(l){2-6} 
 & Terrorism & GTD\textsuperscript{4} & 193 & 34 & 3,520 \\ \bottomrule
\end{tabular}%
}

% \vspace{-1.5em}
\end{table}

\footnotetext[1]{http://snap.stanford.edu/data/loc-gowalla.html}
\footnotetext[2]{http://snap.stanford.edu/data/loc-brightkite.html}
\footnotetext[3]{https://sites.google.com/site/yangdingqi/home}
\footnotetext[4]{https://www.start.umd.edu/gtd/}
\footnotetext[5]{https://www.ntu.edu.sg/home/gaocong/datacode.htm}

% \usepackage{booktabs}
% \usepackage[normalem]{ulem}
\begin{table*}[]
\caption{Performance in Acc@$K$ and MAP on six datasets of LBSN, Terrorism and Transportation domains.} 
\resizebox{\linewidth-6.6cm}{6.5cm}{%
\begin{tabular}{@{}llllllllllll@{}}
\cmidrule(){1-12} 
\multicolumn{6}{c}{Gowalla} &  & \multicolumn{5}{c}{Brightkite} \\
\cmidrule(){1-12} 
 & Acc@1 & Acc@5 & Acc@10 & Acc@20 & MAP &  & Acc@1 & Acc@5 & Acc@10 & Acc@20 & MAP \\
  \cmidrule(r{6pt}){1-6}  \cmidrule(l{3.5pt}){8-12} 
%  \cmidrule(r){1-12} 
TOP & 0.0120 & 0.0414 & 0.0805 & 0.1243 & 0.0338 &  & 0.0824 & 0.2114 & 0.2977 & 0.4168 & 0.1564 \\
U-TOP & 0.1464 & 0.2616 & 0.2695 & 0.2762 & 0.1982 &  & 0.7193 & 0.8223 & 0.8271 & 0.8333 & 0.7703 \\
MF & 0.1347 & 0.2043 & 0.2097 & 0.2156 & 0.1660 &  & 0.7094 & 0.8067 & 0.8105 & 0.8181 & 0.7566 \\
RNN & 0.1051 & 0.2076 & 0.2518 & 0.2937 & 0.1542 &  & 0.7510 & 0.8299 & 0.8537 & 0.8721 & 0.7865 \\
GRU & 0.1090 & 0.2111 & 0.2617 & 0.3112 & 0.1611 &  & 0.7528 & 0.8253 & 0.8474 & 0.8688 & 0.7868 \\
LSTM & 0.1085 & 0.2101 & 0.2585 & 0.3073 & 0.1594 &  & 0.7554 & 0.8283 & 0.8530 & 0.8738 & 0.7889 \\
HST-LSTM & 0.0490 & 0.1194 & 0.1592 & 0.2048 & 0.0883 &  & 0.6532 & 0.8002 & 0.8314 & 0.8562 & 0.7212 \\
STGN & 0.0256 & 0.0784 & 0.1144 & 0.1685 & 0.0590 &  & 0.6435 & 0.7685 & 0.8128 & 0.8605 & 0.7043 \\
STGCN & 0.0424 & 0.1134 & 0.1625 & 0.2249 & 0.0842 &  & 0.6497 & 0.7974 & 0.8287 & 0.8611 & 0.7184 \\
LSTPM & 0.1468 & 0.2506 &  0.2983 &  0.3502 &  0.1998 &  &  0.7554 &  0.8564 &  0.8800 &  0.9057 & 0.8022 \\   \cmidrule(r{6pt}){1-6}  \cmidrule(l{3.5pt}){8-12}
PP-DGAT-Skip & 0.0749 & 0.1366 & 0.1687 & 0.2086 & 0.1096 &  & \textbf{0.7637} & \textbf{0.8709} & \textbf{0.8937} & \textbf{0.9127} & \textbf{0.8128} \\
STP-DGAT & 0.1344 & 0.2414 & 0.2653 & 0.2872 & 0.1856 &  & 0.7338 & 0.8269 & 0.8355 & 0.8470 & 0.7794 \\
STP-UDGAT & \textbf{0.1475} & \textbf{0.2911} & \textbf{0.3285} & \textbf{0.3578} & \textbf{0.2130} &  & 0.7312 & 0.8269 & 0.8355 & 0.8474 & 0.7783 \\
\cmidrule(){1-12} 
\multicolumn{6}{c}{Foursquare-Global} &  & \multicolumn{5}{c}{Foursquare-SG} \\
\cmidrule(){1-12} 
 & Acc@1 & Acc@5 & Acc@10 & Acc@20 & MAP &  & Acc@1 & Acc@5 & Acc@10 & Acc@20 & MAP \\
  \cmidrule(r{6pt}){1-6}  \cmidrule(l{3.5pt}){8-12}
TOP & 0.0118 & 0.0445 & 0.0627 & 0.1048 & 0.0331 &  & 0.0171 & 0.0645 & 0.1056 & 0.1583 & 0.0487 \\
U-TOP & 0.1703 &  0.3231 & 0.3309 & 0.3357 & 0.2367 &  & \textbf{0.0981} &  0.2077 & 0.2601 & 0.3090 & 0.1516 \\
MF & 0.1589 & 0.2626 & 0.2685 & 0.2730 & 0.2043 &  & 0.0723 & 0.1731 & 0.2399 & 0.2960 & 0.1232 \\
RNN & 0.1426 & 0.2896 & 0.3543 & 0.4096 & 0.2119 &  & 0.0207 & 0.0635 & 0.0931 & 0.1255 & 0.0466 \\
GRU & 0.1458 & 0.2861 & 0.3523 & 0.4183 & 0.2157 &  & 0.0145 & 0.0462 & 0.0660 & 0.0905 & 0.0343 \\
LSTM & 0.1445 & 0.2909 & 0.3560 & 0.4191 & 0.2164 &  & 0.0162 & 0.0586 & 0.0814 & 0.1226 & 0.0420 \\
HST-LSTM & 0.0454 & 0.1310 & 0.1876 & 0.2548 & 0.0939 &  & 0.0119 & 0.0355 & 0.0535 & 0.0796 & 0.0288 \\
STGN & 0.0302 & 0.0917 & 0.1528 & 0.2279 & 0.0722 &  & 0.0070 & 0.0249 & 0.0425 & 0.0680 & 0.0222 \\
STGCN & 0.0355 & 0.1233 & 0.1885 & 0.2751 & 0.0874 &  & 0.0090 & 0.0294 & 0.0476 & 0.0753 & 0.0251 \\
LSTPM & 0.1802 & 0.3167 &  0.3795 &  0.4401 &  0.2485 &  & 0.0863 & 0.2032 &  0.2642 &  0.3314 & 0.1465 \\   \cmidrule(r{6pt}){1-6}  \cmidrule(l{3.5pt}){8-12}
PP-DGAT-Skip & 0.1125 & 0.2082 & 0.2498 & 0.2972 & 0.1617 &  & 0.0488 & 0.1187 & 0.1666 & 0.2351 & 0.0909 \\
STP-DGAT & 0.1738 & 0.3310 & 0.3772 & 0.4172 & 0.2476 &  & 0.0949 & 0.2117 & 0.2818 & 0.3574 & 0.1568 \\
STP-UDGAT & \textbf{0.1843} & \textbf{0.3709} & \textbf{0.4359} & \textbf{0.4959} & \textbf{0.2730} & \textbf{} & \textbf{0.0981} & \textbf{0.2155} & \textbf{0.2876} & \textbf{0.3657} & \textbf{0.1604} \\
\cmidrule(){1-12} 
\multicolumn{6}{c}{Transport-SEA} &  & \multicolumn{5}{c}{GTD} \\
\cmidrule(){1-12} 
 & Acc@1 & Acc@5 & Acc@10 & Acc@20 & MAP &  & Acc@1 & Acc@5 & Acc@10 & Acc@20 & MAP \\
  \cmidrule(r{6pt}){1-6}  \cmidrule(l{3.5pt}){8-12}
TOP & 0.0144 & 0.0478 & 0.0700 & 0.1119 & 0.0376 &  & 0.0440 & 0.2589 & 0.4492 & 0.7603 & 0.1716 \\
U-TOP & 0.1448 & 0.2969 & 0.3331 & 0.3388 & 0.2095 &  & 0.7056 & 0.8404 & 0.8673 & 0.9039 & 0.7710 \\
MF & 0.1235 & 0.2755 & 0.3096 & 0.3166 & 0.1860 &  & 0.6864 & 0.8471 & 0.8665 & 0.9193 & 0.7648 \\
RNN & 0.1157 & 0.2634 & 0.3488 & 0.4263 & 0.1899 &  & 0.6702 & 0.8437 & 0.8944 & 0.9528 & 0.7612 \\
GRU & 0.1035 & 0.2378 & 0.3151 & 0.3842 & 0.1723 &  & 0.6805 & 0.8615 & 0.9045 & 0.9536 & 0.7713 \\
LSTM & 0.1116 & 0.2510 & 0.3260 & 0.3984 & 0.1810 &  & 0.7053 & 0.8726 & 0.9037 & 0.9508 & 0.7847 \\
HST-LSTM & 0.0578 & 0.1521 & 0.2141 & 0.2896 & 0.1113 &  & 0.6669 & 0.8371 & 0.8804 & 0.9382 & 0.7503 \\
STGN & 0.0269 & 0.1006 & 0.1556 & 0.2288 & 0.0719 &  & 0.6077 & 0.8017 & 0.8693 & 0.9531 & 0.6944 \\
STGCN & 0.0379 & 0.1228 & 0.1920 & 0.2793 & 0.0896 &  & 0.6612 & 0.8430 & 0.8832 & 0.9448 & 0.7514 \\
LSTPM & 0.1464 &  0.3025 &  0.3891 &  0.4775 &  0.2264 &  &  0.7561 &  0.8925 &  0.9214 &  0.9650 &  0.8214 \\   \cmidrule(r{6pt}){1-6}  \cmidrule(l{3.5pt}){8-12}
PP-DGAT-Skip & 0.1267 & 0.2832 & 0.3637 & 0.4358 & 0.2039 &  & \textbf{0.7695} & \textbf{0.8996} & \textbf{0.9430} & \textbf{0.9741} & \textbf{0.8288} \\ 
STP-DGAT & 0.1472 & \textbf{0.3399} & \textbf{0.4366} & \textbf{0.5391} & \textbf{0.2432} &  & 0.7332 & 0.8579 & 0.9018 & 0.9537 & 0.7948 \\
STP-UDGAT & \textbf{0.1488} & 0.3377 & 0.4207 & 0.5002 & 0.2396 &  & 0.7296 & 0.8676 & 0.9008 & 0.9457 & 0.7961 \\
\cmidrule(){1-12} 
% \cmidrule(r){1-12} 
\end{tabular} %
}

% \vspace{-1.5em}
\end{table*}

Similar to existing works, we use the standard metrics of Acc@$K$ where $K\in\{1,5,10,20\}$ and Mean Average Precision (MAP) for evaluation. Given a test sample, for Acc@$K$, if the ground truth POI is within the top $K$ of the recommendation set, then a score of 1 is awarded, else 0. This helps to understand the performance of the recommendation set up to $K$, whereas MAP scores the quality of the entire recommendation set.

\subsection{Experimental Settings}
We utilise Adam with batch size of 1 using cross entropy loss and ran the experiments with 100 epochs, and set the initial learning rate of 0.001 followed by a decay to 0.0001 at the 10th epoch. We set our POI, user embedding dimension $dim$ and DGAT's $\delta$ to 1,024, and a dropout rate of 0.95. For exploration, we set top $\tau$ new neighbours to 23, $\mu$ and $\beta$ to 5. For RNN, LSTM and GRU, we set the cell state size to 128, same as the recommended size for STGCN. For all other hyper-parameters, we use the same settings as our variants where possible (e.g. POI embedding size $dim$). For all other works, we use their stated recommended settings accordingly.

For HST-LSTM, STGN and STGCN, these models use the \emph{next spatial and temporal intervals} as input to predict $v_{t_{i}}$: i.e. given the visits' details of both next POI $v_{t_{i}}$ and previous POI $v_{t_{i-1}}$, the spatial interval $\Delta d_{t_{i-1}} = d(l_{t_{i}},l_{t_{i-1}}) $ is computed using a distance function $d$ of location coordinates $l_{t_{i}}$ and $l_{t_{i-1}}$ from both visits. The temporal interval $\Delta t_{t_{i-1}} = time_{t_{i}}-time_{t_{i-1}}$ is the difference of timestamps $time$ between both visits. In our experiments, we use the visits of $v_{t_{i-1}}$ and $v_{t_{i-2}}$ to compute $\Delta d_{t_{i-1}}$ and $\Delta t_{t_{i-1}}$ instead of using $v_{t_{i}}$ and $v_{t_{i-1}}$ because the latter requires $v_{t_{i}}$'s visit to be known in advance when the model is trying to predict $v_{t_{i}}$.

\begin{table}[t]
\centering
\caption{Average relative improvement of our best proposed variant over the best baseline method on all datasets from Table 2.}

\resizebox{6.66cm}{0.93cm}{%
\begin{tabular}{lllll}

\toprule
\multicolumn{5}{c}{\textbf{Average Improvement}} \\ 
\midrule
\textbf{Acc@1} & \textbf{Acc@5} & \textbf{Acc@10} & \textbf{Acc@20} & \textbf{MAP} \\ \midrule
1.21\% & 7.45\% & 8.33\% & 6.64\% & 5.32\% \\ \bottomrule
\end{tabular}%
}

\end{table}

\subsection{Results}
We show the comparison results between our proposed variants and baselines from Tables 2 and 3:
\begin{itemize}[leftmargin=*]
  \item From the average relative improvement shown in Table 3, we can conclude that our proposed variants outperform baselines and state-of-the-art LSTPM significantly for all metrics (e.g. highest of 8.33\% for Acc@10).
  \item Looking at each of the six datasets individually, we observe that one of our three variants always has the best results for all metrics.
  \item For Gowalla, Foursquare-Global and Foursquare-SG, we can see our three proposed variants progressively improve performance on all metrics, showcasing the effectiveness of each proposed variant, with STP-UDGAT being the best.
  \item For Brightkite and GTD, our PP-DGAT-Skip variant has the best results. This implies that due to the nature of these two datasets, the exploitation of user's personalized preferences or historical POIs alone is more important to perform well for this recommendation task. 
  \item For Transport-SEA, we observe an interesting trend where STP-UDGAT has the best Acc@1 score and STP-DGAT was the best for the remaining metrics. This suggest that by learning user semantics with UDGAT, this can characterize the user well to perform the best for Acc@1 but in terms of the overall ranked list, learning POI-POI relationships is more important than User-User relationships in the transportation domain. 
  \item U-TOP and LSTPM are the most competitive baselines but did not surpass our variants. For U-TOP, even though it is a simple frequency baseline, it is able to capture the human mobility behaviours well as users would simply tend to visit their most frequent POIs. Comparing STP-UDGAT to LSTPM, one of the several key differences is in our proposed inclusion of the exploration module to also consider new unvisited POIs that are still relevant to the user, as well as our proposed explore-exploit architecture to balance the trade-offs. 
  
  \item HST-LSTM, STGN and STGCN do not perform as well, partly as they rely heavily on spatial and temporal intervals between $v_{t_{i}}$ and $v_{t_{i-1}}$ and were not robust to learn from intervals between $v_{t_{i-1}}$ and $v_{t_{i-2}}$ for their works.
  \item Only for Foursquare-SG, our best performing STP-UDGAT has the same Acc@1 score as U-TOP, but was best for the remaining metrics. This is likely due to a high dropout rate used in STP-UDGAT to prevent overfitting to the most frequent POI, therefore resulting to a notable trend of increasing improvements from Acc@1 towards Acc@20 when comparing U-TOP and STP-UDGAT.
  
\end{itemize}

\subsection{Performance for Cold Start Problem}
To ensure robustness to little training data, we do a separate preprocessing of keeping POIs visited by more than 1 user and keeping users with visit counts less than 10 to simulate the cold start scenario. We evaluate the cold start recommendation performance on the Foursquare-Global dataset for Acc@1, our largest dataset with worldwide POIs. Fig. 6 shows STP-UDGAT surpassing all baselines and state-of-the-art LSTPM on test set, demonstrating better performances even with short POI visit sequences.

\subsection{Ablation Study}
In this section, we perform two sets of ablation studies of STP-UDGAT and its explore-exploit performances. Same as the cold start problem, we perform the analysis on our largest dataset of Foursquare-Global with POIs across 40 countries for Acc@1.

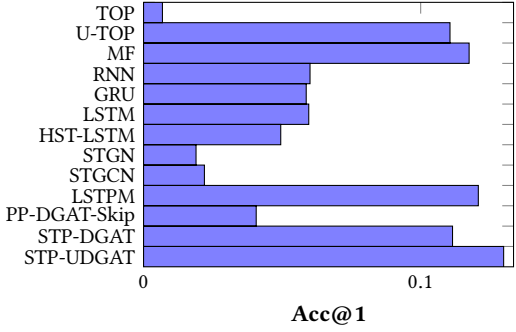
\begin{figure}[t]
\hspace*{-1.5cm} % good for manual moving
\centering
\begin{tikzpicture} % cold STart
\begin{axis}[
  width=6.5cm
    ,height=5.1cm,
  xmin=0,xmax=0.1335,
% ylabel=Age,
  xlabel=\textbf{Acc@1},
  enlargelimits=false,
  ytick={0,1,...,12},
% yticklabel style={
%    xshift=10},
  yticklabels={TOP,U-TOP,MF,RNN,GRU,LSTM,HST-LSTM,STGN,STGCN,LSTPM,PP-DGAT-Skip,STP-DGAT,STP-UDGAT},
  xtick distance=0.1,
% yticklabel interval boundaries,
% xbar ,
  y dir=reverse, %change to ascending orer
% axis lines=left,
% xlabel shift={-10pt},
% xbar interval,
% ylabel near ticks,
    xlabel near ticks,
  tick label style={font=\small}  ,
  yticklabel style={xshift=0.0cm,yshift=-0.14cm}, % IMPORTANT, otherwise bar chart text won't shift correctly.
  xticklabel style={/pgf/number format/fixed} % disable scientific notation
]
\addplot[xbar interval,fill=blue!50!white]
  coordinates {(0.00680957615732258,0) (0.110571637962623,1) (0.117472822767802,2) (0.0600647367778184,3) (0.0586906070599731,4) (0.0596066935385367,5) (0.0494992060583852,6) (0.0189324538903139,7) (0.0219555392695737,8) (0.120792865432777,9) (0.0406437034322706,10)
  (0.111487724441187,11)
  (0.129931598876267,12)
  (0.129931598876267,13)

    };
\end{axis}
%\caption{Performance of Cold Start}
\end{tikzpicture}
%\raggedbottom
%   \vspace*{-0.2cm}
\vspace*{-0.45cm}
\caption{Cold Start Performance on Foursquare-Global.} \label{fig:M1}
% \vspace*{-0.2cm}
%\raggedbottom

\end{figure}

\begin{figure}[t]
% to adjust spacing on top and below caption
\captionsetup[subfigure]{aboveskip=1.2pt,belowskip=9pt}
%\centering
  \begin{subfigure}[t]{\linewidth} \hspace{13pt}
%    \centering
\begin{tikzpicture} %U-GAT
    \begin{axis}
        [
        ,width=6.5cm
        ,height=2.8cm
%        ,xlabel=Test
        ,ylabel=\textbf{Acc@1}
        ,y tick label style={
        /pgf/number format/.cd,
            %fixed,
            %fixed zerofill,
            precision=3,
        /tikz/.cd
    }
    ,ytick distance=0.01
    %,ytick = {0.17, 0.18,0.19}
        %,xtick=data,
        ,xtick={0,1,...,2}
        ,xticklabels={STP-DGAT,STP-DGAT-Embed,STP-UDGAT}
%      ,xmin=0, xmax=100
      ,ymin=0.1685, ymax=0.189
%   ,ymajorgrids=true
%   ,grid style=dashed
    ,label style={font=\small},
        tick label style={font=\small}  
        ]
        \addplot+[sharp plot,mark=square] coordinates
        {(0,0.1738251539399164) (1,0.17286204554552473) (2,0.18434456903569893)};
        % {(0,0.174958507063918) (1,0.173510752180346) (2,0.182399686048027)};
    \end{axis}
\end{tikzpicture}
%\hspace*{-30mm}
    \caption{Performance of UDGAT.}
  \end{subfigure}
%  \hfill
%   \begin{subfigure}[t]{\linewidth}\hspace{13pt}
% %    \centering
%  \begin{tikzpicture} % num neigh
%     \begin{axis}
%         [
%         ,width=6.2cm
%         ,height=2.93cm
% %        ,xlabel=$K$
%         ,ylabel=\textbf{Acc@1}
%         ,y tick label style={
%         /pgf/number format/.cd,
%             %fixed,
%             %fixed zerofill,
%             precision=3,
%         /tikz/.cd
%     }
%     ,ytick distance=0.003
%     %,ytick = {0.17, 0.18,0.19}
%         %,xtick=data,
%         ,xtick={1,2,...,10}
% %        ,xticklabels={STP-GAT,STP-GAT-Embed,STP-UDGAT}
% %      ,xmin=0, xmax=100
%       ,ymin=0.176, ymax=0.181
% %   ,ymajorgrids=true
% %   ,grid style=dashed
%     ,label style={font=\footnotesize},
%         tick label style={font=\footnotesize}  
%         ]
%         \addplot+[sharp plot,mark=square] coordinates
%         {(1,0.1769786170820451) (2,0.17706205274213138) (3,0.17845549573376066)
%         (4,0.17774370634245867) (5,0.17716775290450468) (6,0.17877507005446847)
%         (7,0.1769512924656017)
%         (8,0.1801689628340229)
%         (9,0.17929142659236091)
%         (10,0.18032087870568947)
%         };
%     \end{axis}
% \end{tikzpicture}
%     \caption{Number of new neighbours $\tau$}
%   \end{subfigure}
%\hfill
%  \medskip

  \begin{subfigure}[t]{\linewidth}\hspace{13pt}
%    \centering
    \begin{tikzpicture} % proposed graphs
    \begin{axis}
        [
        ,width=6.5cm
        ,height=2.8cm
%        ,xlabel=Graph Type
%     ,xlabel=
        ,ylabel=\textbf{Acc@1}
        ,y tick label style={
        /pgf/number format/.cd,
            %fixed,
            %fixed zerofill,
            precision=3,
        /tikz/.cd
    }
    ,ytick distance=0.005
    %,ytick = {0.17, 0.18,0.19}
        %,xtick=data,
        ,xtick={1,2,...,7}
        ,xticklabels={S,T,P,STP}
%      ,xmin=0, xmax=100
      ,ymin=0.177, ymax=0.186
%   ,ymajorgrids=true
%   ,grid style=dashed
    ,label style={font=\small},
        tick label style={font=\small}  
        ]
        \addplot+[sharp plot,mark=square*] coordinates
        {(1,0.1790276259743523) (2,0.1823234469947412) (3,0.1818046166232613)
        (4,0.18434456903569893) 
        };
    \end{axis}
\end{tikzpicture}
% \vspace{-0.5\baselineskip}
\caption{Effectiveness of proposed graphs.}
  \end{subfigure}
%  \hfill
  \begin{subfigure}[t]{\linewidth}\hspace{13pt}
%    \centering
    \begin{tikzpicture} % walks vs adjancy
    \begin{axis}
        [
        ,width=6.5cm
        ,height=2.8cm
%        ,xlabel=Masked Attention Type
        ,ylabel=\textbf{Acc@1}
        ,y tick label style={
        /pgf/number format/.cd,
            %fixed,
            %fixed zerofill,
            precision=3,
        /tikz/.cd
    }
    ,ytick distance=0.005
    %,ytick = {0.17, 0.18,0.19}
        %,xtick=data,
        ,xtick={1,2,...,7}
        ,xticklabels={$A$,$RW$,$A+RW$}
%      ,xmin=0, xmax=100
      ,ymin=0.176, ymax=0.187
%   ,ymajorgrids=true
%   ,grid style=dashed
    ,label style={font=\small},
        tick label style={font=\small}  
        ]
        \addplot+[sharp plot,mark=*] coordinates
        {(1,0.17872424401894424) (2,0.1789313713584037) (3,0.18434456903569893)
        };
    \end{axis}
\end{tikzpicture}
    \caption{Performance of masked self-attention options.}
  \end{subfigure}
    \begin{subfigure}[t]{\linewidth}\hspace{13pt}
%    \centering
    \begin{tikzpicture} % walks vs adjancy
    \begin{axis}
        [
        ,width=6.5cm
        ,height=2.8cm
%        ,xlabel=Masked Attention Type
        ,ylabel=\textbf{Acc@1}
        ,y tick label style={
        /pgf/number format/.cd,
            %fixed,
            %fixed zerofill,
            precision=3,
        /tikz/.cd
    }
    ,ytick distance=0.008
    %,ytick = {0.17, 0.18,0.19}
        %,xtick=data,
        ,xtick={1,2,...,7}
        ,xticklabels={Scalar,Dimensional}
%      ,xmin=0, xmax=100
      ,ymin=0.175, ymax=0.187
%   ,ymajorgrids=true
%   ,grid style=dashed
    ,label style={font=\small},
        tick label style={font=\small}  
        ]
        \addplot+[sharp plot,mark=diamond*] coordinates
        {(1,0.177080494047056) (2,0.18434456903569893)
        };
    \end{axis}
\end{tikzpicture}
    \caption{Effectiveness of DGAT's dimensional attention.}
  \end{subfigure}
%     \begin{subfigure}[t]{\linewidth}\hspace{13pt}
% %    \centering
%     \begin{tikzpicture} % walks vs adjancy
%     \begin{axis}
%         [
%         ,width=6.2cm
%         ,height=2.93cm
% %        ,xlabel=Masked Attention Type
%         ,ylabel=\textbf{Acc@1}
%         ,y tick label style={
%         /pgf/number format/.cd,
%             %fixed,
%             %fixed zerofill,
%             precision=3,
%         /tikz/.cd
%     }
%     ,ytick distance=0.02
%     %,ytick = {0.17, 0.18,0.19}
%         %,xtick=data,
%         ,xtick={1,2,...,7}
%         ,xticklabels={Exploit,Explore,Explore-Exploit}
% %      ,xmin=0, xmax=100
%       ,ymin=0.145, ymax=0.193
% %   ,ymajorgrids=true
% %   ,grid style=dashed
%     ,label style={font=\footnotesize},
%         tick label style={font=\footnotesize}  
%         ]
%         \addplot+[sharp plot,mark=triangle*] coordinates
%         {(1,0.15249328691521905) 
%         (2,0.15920108668762623) (3,0.18434456903569893)
%         };
%     \end{axis}
% \end{tikzpicture}
%     \caption{Impact of explore-exploit}
%   \end{subfigure}
  \vspace*{-0.45cm}
  \caption{Analysis of STP-UDGAT on Foursquare-Global.}
\end{figure}
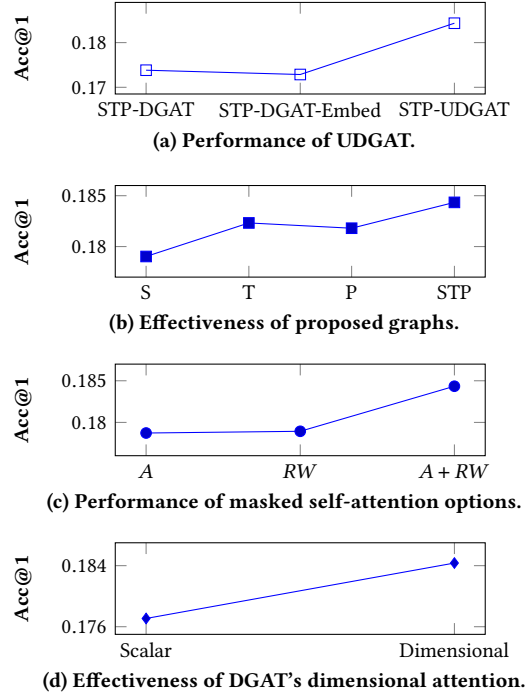

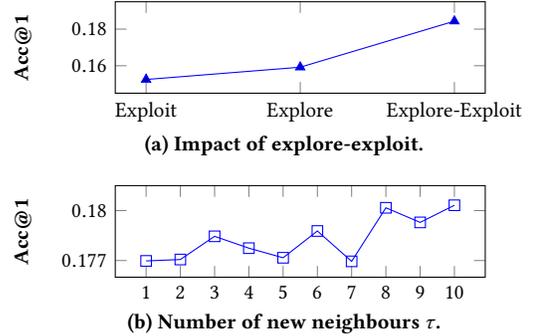
\begin{figure}[t]
% to adjust spacing on top and below caption
\captionsetup[subfigure]{aboveskip=1.2pt,belowskip=9pt}
%\centering
    \begin{subfigure}[t]{\linewidth}\hspace{13pt}
%    \centering
    \begin{tikzpicture} % walks vs adjancy
    \begin{axis}
        [
        width=6.5cm
        ,height=2.8cm
%        ,xlabel=Masked Attention Type
        ,ylabel=\textbf{Acc@1}
        ,y tick label style={
        /pgf/number format/.cd,
            %fixed,
            %fixed zerofill,
            precision=3,
        /tikz/.cd
    }
    ,ytick distance=0.02
    %,ytick = {0.17, 0.18,0.19}
        %,xtick=data,
        ,xtick={1,2,...,7}
        ,xticklabels={Exploit,Explore,Explore-Exploit}
%      ,xmin=0, xmax=100
      ,ymin=0.145, ymax=0.195
%   ,ymajorgrids=true
%   ,grid style=dashed
    ,label style={font=\small},
        tick label style={font=\small}  
        ]
        \addplot+[sharp plot,mark=triangle*] coordinates
        {(1,0.15249328691521905) 
        (2,0.15920108668762623) (3,0.18434456903569893)
        };
    \end{axis}
\end{tikzpicture}
    \caption{Impact of explore-exploit.}
  \end{subfigure}
  \begin{subfigure}[t]{\linewidth}\hspace{13pt}
%    \centering
 \begin{tikzpicture} % num neigh
    \begin{axis}
        [
        width=6.5cm
        ,height=2.8cm
%        ,xlabel=$K$
        ,ylabel=\textbf{Acc@1}
        ,y tick label style={
        /pgf/number format/.cd,
            %fixed,
            %fixed zerofill,
            precision=3,
        /tikz/.cd
    }
    ,ytick distance=0.003
    %,ytick = {0.17, 0.18,0.19}
        %,xtick=data,
        ,xtick={1,2,...,10}
%        ,xticklabels={STP-DGAT,STP-DGAT-Embed,STP-UDGAT}
%      ,xmin=0, xmax=100
      ,ymin=0.176, ymax=0.1815
%   ,ymajorgrids=true
%   ,grid style=dashed
    ,label style={font=\small},
        tick label style={font=\small}  
        ]
        \addplot+[sharp plot,mark=square] coordinates
        {(1,0.1769786170820451) (2,0.17706205274213138) (3,0.17845549573376066)
        (4,0.17774370634245867) (5,0.17716775290450468) (6,0.17877507005446847)
        (7,0.1769512924656017)
        (8,0.1801689628340229)
        (9,0.17929142659236091)
        (10,0.18032087870568947)
        };
    \end{axis}
\end{tikzpicture}
    \caption{Number of new neighbours $\tau$.}
    
  \end{subfigure}
%\hfill
%  \medskip

%     \begin{subfigure}[t]{\linewidth}\hspace{13pt}
% %    \centering
%     \begin{tikzpicture} % walks vs adjancy
%     \begin{axis}
%         [
%         ,width=6.2cm
%         ,height=2.93cm
% %        ,xlabel=Masked Attention Type
%         ,ylabel=\textbf{Acc@1}
%         ,y tick label style={
%         /pgf/number format/.cd,
%             %fixed,
%             %fixed zerofill,
%             precision=3,
%         /tikz/.cd
%     }
%     ,ytick distance=0.02
%     %,ytick = {0.17, 0.18,0.19}
%         %,xtick=data,
%         ,xtick={1,2,...,7}
%         ,xticklabels={Exploit,Explore,Explore-Exploit}
% %      ,xmin=0, xmax=100
%       ,ymin=0.145, ymax=0.193
% %   ,ymajorgrids=true
% %   ,grid style=dashed
%     ,label style={font=\footnotesize},
%         tick label style={font=\footnotesize}  
%         ]
%         \addplot+[sharp plot,mark=triangle*] coordinates
%         {(1,0.15249328691521905) 
%         (2,0.15920108668762623) (3,0.18434456903569893)
%         };
%     \end{axis}
% \end{tikzpicture}
%     \caption{Impact of explore-exploit}
%   \end{subfigure}
  \vspace*{-0.45cm}
  \caption{Analysis of explore-exploit on Foursquare-Global.}
\end{figure}

\paragraph{\textbf{STP-UDGAT}} Fig. 7 shows the ablation analysis for STP-UDGAT where various components were deactivated:
\begin{itemize}[leftmargin=*]
\item Fig. 7(a) shows STP-UDGAT surpassing STP-DGAT-Embed, where the latter concatenates a user embedding directly instead of UDGAT's representation in Eq. (14). Also, STP-DGAT performed better than STP-DGAT-Embed, indicating that the direct inclusion of user embedding does not always help the task. In contrast, STP-UDGAT has a significant increase of performance.
% due to the ability to reduce overfitting by allowing users to learn to attend to other users through the proposed UDGAT.
% Results show that for this challenging dataset, the inclusion of user embedding alone performed worse as can be seen by the difference of STP-DGAT and STP-DGAT-Embed.
\item Fig. 7(b) illustrates the usage of the STP graphs individually, achieving sub-optimal performance, whereas they performed best when combined together, showing the effectiveness of our proposed STP graphs.

\item Fig. 7(c) shows better performance of our newly proposed random walk masked self-attention option $RW$ over GAT's classical adjacency option $A$ for the exploration module. In addition, the best result is achieved when $A$ and $RW$ are used together.

\item Fig. 7(d) demonstrates the effectiveness of DGAT where a large increase of performance can be seen for dimensional attention, in comparison with the scalar attention used in classical GAT.

\end{itemize}

\paragraph{\textbf{Explore-Exploit}} Fig. 8 illustrates the ablation analysis of the explore-exploit component of our STP-UDGAT model. Fig. 8(a) shows three scenarios of exploit only, explore only and explore-exploit. Based on the illustration of STP-UDGAT in Fig. 5: 
\begin{itemize}[leftmargin=*]
\item Exploitation only deactivates the exploration module, where learning of new unvisited POIs are not considered.

\item Exploration only deactivates PP-DGAT or the exploitation module, where learning from user's PP graph or historical POIs are not considered.
\item Explore-Exploit is the proposed STP-UDGAT model that considers both explore and exploit by learning the balance.
\end{itemize}

Fig. 8(a) shows that our proposed exploration module performs better than exploitation of users' historical POIs, indicating that our newly identified POIs via global STP graphs are indeed relevant to the users and benefit learning even when they have never visited these POIs before. A large increase can also be seen when combining both to perform explore-exploit, demonstrating that STP-UDGAT is able to balance the trade-offs by learning optimal parameters. Additionally, Fig. 8(b) shows an overall increasing trend of performance based on increasing $\tau$ (newly explored unvisited POIs).

\begin{figure}[t] 
  \centering

  \includegraphics[width=\columnwidth,height=5.3  cm]{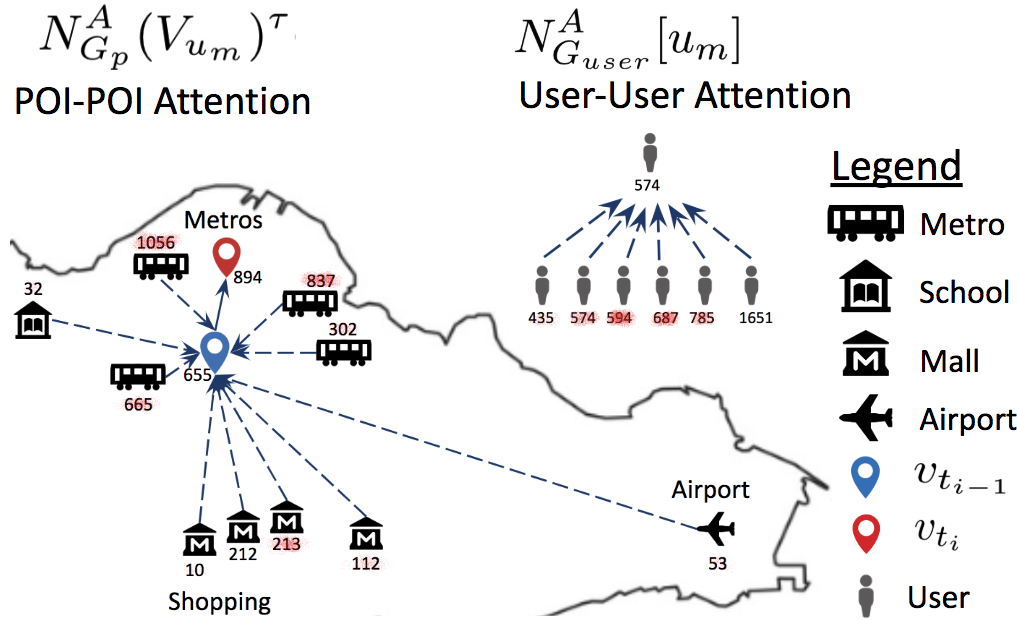}

  \caption{Sample prediction from Foursquare-SG dataset with STP-UDGAT and its computed attention weights (red).}

\end{figure}

\subsection{Case Study: Interpretability}
Each of STP-UDGAT's eight DGAT layers is interpretable. For instance, given a test sample of POI 655 to try to predict POI 894 (both metros) for user 574, Fig. 9 shows the legend, a DGAT layer of newly explored POIs $N^{A}_{G_{p}}(V_{u_{m}})^{\tau}$ from the preference graph $G_{p}$ for POI-POI attention and another DGAT layer of $N^{A}_{G_{user}}[u_{m}]$ for user-user attention. We observe that STP-UDGAT computes higher coefficients to mostly nearby metros, over distant malls and airport to try to predict POI 894, a metro. We can also see user 574 attending more to users 594, 687 and 785 than herself. These validates the goal of STP-UDGAT, supporting interpretability and model transparency as compared to existing RNN models.

\section{Conclusion}
This paper proposed a novel explore-exploit STP-UDGAT model for the next POI recommendation task. Experimental results on six real-world datasets prove the effectiveness of the proposed approach for multiple applications including LBSN, transport and terrorism. For future work, we aim to study how pick-up points in transportation domain can help support the recommendation task.

\section*{Acknowledgment}
This work was funded by the Grab-NUS AI Lab, a joint collaboration between GrabTaxi Holdings Pte. Ltd. and National University of Singapore, and the Industrial Postgraduate Program (Grant: S18-1198-IPP-II) funded by the Economic Development Board of Singapore.

\bibliographystyle{ACM-Reference-Format}
\bibliography{sample-base}

\end{document}